# Shaped pulse electric-field construction and interferometric characterization: The SPECIFIC method

Matthew A. Coughlan[1], Mateusz Plewicki[1], Stefan M. Weber[2], Pamela Bowlan[3], Rick Trebino[3], and Robert J Levis[1*]

[1]Department of Chemistry, Center for Advanced Photonics Research, Temple University, Philadelphia, Pennsylvania 19122, USA
[2] GAP – Biophotonics, Université de Genève, Rue de l'École-de-Médecine 20, CH-1211 Genève 4
[3]Georgia Institute of Technology, School of Physics, 837 State St NW, Atlanta, GA 30332 USA
*Corresponding author: rjlevis@temple.edu

**Abstract:** A method is reported for creating, generating, and measuring parametrically shaped pulses for time-bandwidth product >>5, which consists of a parametric pulse-shaping algorithm, a spatial light modulation system and a single shot interferometric characterization scheme (SEA TADPOLE) . The utilization of these tools marks the inception of a new method called SPECIFIC, shaped-pulse electric-field construction and interferometric characterization, capable of producing complex shaped laser pulses for coherent control experiments.

---

## 1. Introduction:

In recent years, manipulating the spectral phase and amplitude of ultrafast laser pulses has enabled chemists and physicists unprecedented ability to control non-linear systems [1-5]. In most of these experiments, the optimal pulses are generated by a search algorithm that results in complex pulse shapes where the complexity arises from modulation of hundreds to thousands of distinct spectral phases and amplitudes. The complexity arising from such modulation can be characterized by a large time bandwidth product (TBP), defined as the product of the root mean square of the temporal width and the root mean square of the spectral width of a laser pulse[6]. Therefore, the ability to synthesize and characterize such complex laser pulses is of paramount concern for the coherent-control community.

Femtosecond pulse shaping technology has been employed in many spectroscopic applications. For instance, shaping has been used as a means to enhance resonant versus nonresonant contribution to Raman signals or as a tool to extract information regarding a system of interest [7-11]. Femtosecond single beam CARS is now possible using pulse shaping technology to measure linewidths comparable to conventional CARS experiments with picojoule pulse energies [12]. Using femtosecond pulse shaping, vibrational modes in the Raman spectrum of β-carotene can be selectively excited [13]. Tailored multipulse sequences with temporal spacing on the order of the period of the vibrational mode of interest, can produce Raman spectra filtered from unwanted excitation [14]. From a spectroscopic standpoint, there is a need to generate temporal profiles that can be created, and verified on the fly according to a specified design.

The main principles underlying laser pulse shaping take root in electrical engineering, particularly linear filtering. Linear filtering is a method used for processing electrical signals and utilizes familiar components such as resistors, capacitors, and inducers to produce shaped electrical waveforms. In the frequency domain linear filtering can be represented by the following mathematical relations. First there is the frequency response $H(\omega)$ which is related to the linear filter output through equation 1.

$$E_{out}(\omega) = E_{in}(\omega)H(\omega) \qquad (1)$$

Where $E_{out}(\omega)$ and $E_{in}(\omega)$ are the output and inputs of the linear filter, respectively. In the time domain the linear filter can be represented by the Fourier transform pairs in equation 2.

$$H(\omega) = \int dt h(t) e^{-i\omega t} \qquad (2)$$

where $h(t)$ is represented by equation 3,

$$h(t) = \frac{1}{2\pi} \int d\omega H(\omega) e^{i\omega t} \qquad (3)$$

Because of the Fourier relation of equations 2 and 3, a shaped optical waveform can be generated in the temporal domain by introducing a phase and amplitude filter in the spectral domain[15]. Here we demonstrate a frequency domain approach to synthesize a required temporal sequence of pulses, each with a desired energy and phase profile. The design method evaluates the user-defined temporal pulse structure taking into account such parameters as the number of desired pulses, their relative energy, and temporal phase. To accomplish the design, a series of temporally separated sub-pulses are generated in frequency space, the spectral phases (expanded in Taylor series up to fourth order) and amplitudes are then super-imposed to obtain the appropriate pulse shaping mask to produce the desired temporal pulse structure. The parametrically designed pulses are then experimentally generated and verified.

Parametric pulse shaping is important when a model resulting in a theoretically predicted shape can be verified by experimental generation and characterization of the pulse shape. The key to parametric shaping, then, is the generation of a pulse shape from a theoretical prediction with high fidelity. Previously reported pulse shaping algorithms provide the ability to implement a specified waveform in the time domain by calculating the necessary spectral amplitudes and phases for the laser pulse [16-20], in general for the case of small TBP, simple pulses. The paramount objective for parametric pulse shaping is that the generated pulse resembles the theoretical pulse. Therefore, the similarity of the calculated and measured temporal amplitudes is the indicator of success of the parametric pulse shape experiment. The components of the parametric shape in terms of sign and shape of temporal features

(positive/negative, linear, quadratic, cubic chirp, etc.), then qualify as an indicator of success. Measuring all of these components for a designed pulse shape represents a rigorous method of characterizing and determining the success of a parametric pulse shaping system. Pulse design, synthesis and measurement techniques, sensitive to the spectral phase and amplitude, as wells as the temporal envelope and phase, are necessity for quantitative parametric shaping experiments.

Most methods for measuring ultrashort laser pulses fail for complex pulses. The oldest technique for measuring pulses is autocorrelation [6]. Unfortunately, autocorrelation and the interferometric version yield no information regarding the actual temporal intensity and phase of a pulse and so have almost no utility for the pulse shaping community.

The most popular pulse-measurement instrument for measuring temporal or spectral amplitude and phase is frequency-resolved optical gating (FROG) [6]. The FROG setup is similar to non-collinear autocorrelation, with the addition of a spectrometer for detection. FROG can provide phase and amplitude information for simple pulses. However, for complex pulses (TBP > 10), the FROG iterative reconstruction is time-consuming (seconds) and, depending on which version of FROG is used, results in an convergence for such complex pulses only 90 to 95% of the time [20]. Consequently, FROG works very well for moderately complex pulses but becomes increasingly inconvenient as the complexity of the pulses increases beyond a TBP larger than approximately ten [20].

A version of FROG, called XFROG, which uses a well characterized reference pulse, very reliably measures even extremely complicated pulses, with TBP up to 100 (and probably higher) [21]. But XFROG, like FROG, takes some time for its algorithm to converge. In addition, its non-linear optical interaction implies reduced sensitivity compared to linear methods.

Another method for measuring shaped pulses, called MIIPS, was recently introduced[22], which involves using the shaper to add particular spectral-phase functions to the pulse and then measuring the second-harmonic spectrum as a function of the applied spectral phase. Thus far, MIIPS has been used only to measure simple pulses with minimal structure [23]. Significant drawbacks to MIIPS, are that the method is inherently multi-shot and so requires time to generate a data trace, and it is also a nonlinear technique. The method also requires a stable input pulse train. Finally, MIIPS does not constitute an *independent* measure of the pulse shaper's performance because the same shaper that generates the pulse is also an essential component in the measurement of the pulse.

Spectral interferometry (SI) is in principal the easiest, fastest, most reliable, and most sensitive technique because it is linear-optical, single-shot, and independent. Moreover, the measured trace can be directly inverted to reconstruct the pulse [24, 25]. However, traditional SI is a difficult experiment to implement, because the method requires precisely-aligned, interferometrically stable collinear beams, stable mounting of optical components, and stringent spatial mode matching. If any of these requirements is not satisfied, the interference fringes degrade (or disappear), and the error of the measurement increases significantly. Additionally, the most reliable reconstruction method for SI involves Fourier filtering the fringes along the frequency (or time) axis, and this process results in a large reduction of spectral resolution,[24] usually by a factor of five.

A nonlinear-optical version of SI, called spectral shearing interferometry for direct electric field reconstruction (SPIDER), can, in principle, provide spectral phase and amplitude [26]. Because SPIDER is nonlinear-optical, SPIDER is not as sensitive as a linear method. Finally, the method has extremely stringent calibration requirements with no feedback related to measurement accuracy [27]. SPIDER and its variations are typically used for measuring ultrashort pulses.

Another variation of SI, called SEA TADPOLE or Spatially Encoded Arrangement for Temporal Analysis by Dispersing a Pair of Light E-Fields, has recently been demonstrated as a simple and robust variation of spectral interferometry for laser pulse shape characterization

[28]. The advantages of SEA TADPOLE over traditional SI originate from the use of optical fibers, and the two-dimensional interferogram which is made using temporally overlapping and crossing beams [29-34]. The optical fibers perform the essential functions of desensitizing the device to optical and laser instabilities, and ensure that the spatial modes of the interfering beams will be identical and overlap spatially. The 2-D fringes allow for recovery of the spectral phase with high spectral resolution as a result of a zero temporal delay and Fourier filtering along the *spatial* axis of the camera rather than the frequency axis [29]. SEA TADPOLE has been used to measure very complex pulse shapes typically generated in an optimal control experiment [35] with TBP as high as several hundred. SEA TADPOLE recovered phase information that was discontinuous (i.e. had large phase jumps) and therefore can reliably characterize the highly complex pulses expected from pulse shapers. As a result, we have chosen SEA TADPOLE for the measurement of shaped pulses in our apparatus. Other researchers using other methods of pulse shaping have shown that SEA TAPDOLE is a useful tool for characterizing shaped pulses [36, 37, 38] (though they do not use fiber optics in their experimental setup).

Here, for the first time we combine an algorithm for complex parametric pulse generation with a laser pulse shaping system and SEA TADPOLE to provide a new parametric pulse shaping and pulse shape confirmation apparatus, which we call the SPECIFIC (shaped-pulse electric-field construction and interferometric characterization). Using the SPECIFIC method we demonstrate a systematic and accurate way to engineer laser pulses by calculating, applying, and measuring a series of desired pulse shapes by comparing the measured and specified temporal phase and intensity.

**2. Experimental:**

We performed measurements to demonstrate the accuracy and precision of the SPECIFIC method for pulse shaping using parametric pulse shaping of the output of a KM:Labs Ti:Sapphire oscillator in conjunction with SEA TADPOLE interferometric analysis. The parametric pulse masks were applied to a CRI (Cambridge Research Instruments) SLM-2 X 128 spatial light modulator in reflective geometry. The reflective geometry consists of a grating (1200grooves/mm), a cylindrical mirror (f=210mm) and a pulse shaper which has a high reflective dielectric mirror placed after the second array of the spatial light modulator. The spatial light modulator is tilted down at a slight angle to allow the laser beam to propagate below the incoming beam. The SEA TADPOLE experimental apparatus[39] consists of a two equal length single mode fibers, one for the reference and one for the unknown pulse. The beams emerging from the fibers cross at a small angle after being collimated by a spherical lens placed a focal length away from the fibers' ends. A CCD camera is placed at the beams' crossing point in order to record their interference.

We map wavelength to the horizontal dimension of the camera using a diffraction grating and a cylindrical lens and in the vertical dimension the pulses cross. The pulse shaper was placed in one arm of the interferometer. The other arm of the device contains a delay line to compensate for the distance traveled through the pulse shaper so that the interfering pulses temporally overlap at the camera.

To reconstruct the shaped pulse's electric field from the interferogram we use a standard Fourier filtering algorithm that has been discussed in detail in previous papers [28, 31]. This inversion algorithm is potentially very fast (video rate speed or better). All other experimental details about SEA TADPOLE can be found in these references [28, 35].

*2.1 Parametric Pulse Generation*:

The principal of parametric pulse generation comes from the use of a "filter function," $H(\omega)$, which is a complex quantity satisfying the relation of Equation 1. In the filter function $E_{out}(\omega)$ and $E_{in}(\omega)$ describe the desired output and input pulse in spectral domain. Thus, the filter function is directly calculated and decomposed into the real component and complex part, where $R(\omega)$ describes the amplitude and $\phi(\omega)$ corresponds to the phase filter.

$$\tilde{E}_{in}(\omega)/\tilde{E}_{out}(\omega) = \tilde{H}(\omega) = R(\omega) e^{i\phi(\omega)} \quad (4)$$

An electric field consisting of any number of sub pulses can therefore be written as

$$\tilde{E}_{out}(\omega) = \tilde{E}_{in}(\omega)\tilde{H}_1(\omega) + \tilde{E}_{in}(\omega)\tilde{H}_2(\omega) + ... \\ + \tilde{E}_{in}(\omega)\tilde{H}_N(\omega) = \tilde{E}_{in}(\omega)\sum_{i=N}^{0}\tilde{H}_i(\omega) \quad (5)$$

where the $H_i(\omega)$ characterizes the individual, complex sub pulse filter functions. These filter functions can be simply added in order to generate a train of pulses. The algorithm based on the filter function is depicted in Fig. 1.

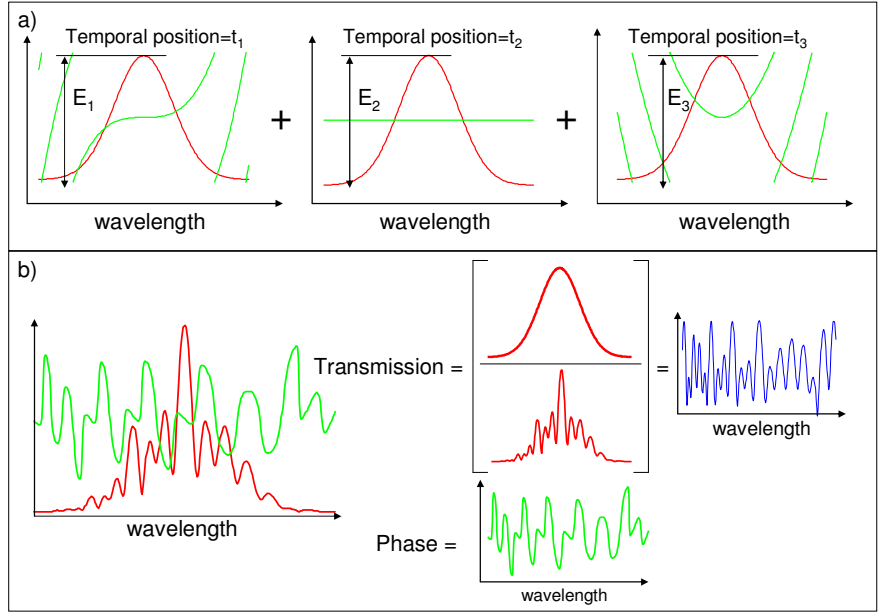

Fig. 1. Schematic description of the parametric construction of the desired sub-pulses. The algorithm is based on: a), specifying the temporal separation and the spectral phases of the pulses; and b), generating the phase and transmission filters.

The number of sub-pulses in the pulse desired pulse shape are first defined with specific temporal positions and corresponding relative energies. The first temporal position is chosen as the first parameter and is translated into the spectral domain by adding a linear phase ramp defined as $\varphi_l(\omega) = \omega*T$. This can be regarded as a first order phase, where $T$ is the temporal position. Next, the desired higher order spectral phases (GVD, TOD and etc.) are included for the first sub-pulse. The remaining sub-pulses are then treated in the same manner, defining a temporal shift and higher order phase functions as desired. The complex electric fields of the sub pulses are superimposed in the spectral domain, thus creating an elaborate and often non-intuitive interference containing real and imaginary components. The absolute and argument of the complex interference directly specifies the modulated spectral amplitude and phase.

Given the specified spectral amplitude and phase, the filter function to be applied to the shaper, $H(\omega)$, can be calculated to produce the desired pulses. The amplitude modulation, $R(\omega)$, is given by ratio of the calculated and input spectra, whereas the phase filter $\phi(\omega)$ is simply the difference between the input field phase and that specified. In the case of a transform limited pulse as an input where the initial spectral phase is constant, the phase filter becomes the phase obtained from the interference of the all sub pulses, as shown in Fig. 1. In such situation the measurement of the pulse structure becomes equivalent to measuring the filter function. The advantage of calculating the parameters in the spectral domain involves the intuitive picture of each sub-pulse with regard to higher order phase, timing and spectral content. These are the parameters that form the natural basis for physical interpretation, dispersion management in optical calculations, and human insight into photochemical mechanism, and thus offers a valuable perspective for parametric pulse shaping.

*2.2 Data:*

As a first example we design, synthesize, and measure a two pulse structure with flat phase and a separation of 400fs as shown in Fig. 2(a).

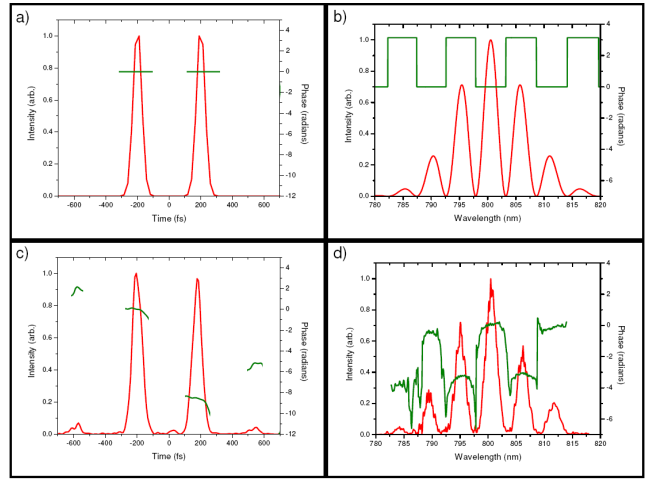

Fig. 2. a), The specified temporal amplitude and phase for a two pulse sequence. b), The target pulse shape's spectral phase (green) and intensity (red). c), The measured temporal intensity and phase from the pulse sequence generated by the SPECIFIC algorithm as measured by SEA TADPOLE. d), The measured spectral phase (green) and amplitude (red) of the target pulse shape.

Applying the SPECIFIC algorithm we calculate the required phase and spectral intensity profile that must be generated by the spatial light modulator as shown in Fig. 2(b). The calculated spectral phase of Fig. 2(b) is a periodic step function, which varies by π phase steps. The phase across each spectral fringe is flat.

The temporal profile measured by SEA TADPOLE shown in Fig. 2(c) displays high correlation to the desired time-dependent pulse shape (Fig. 2(a)) in both intensity and phase. While the profile of the calculated and recovered temporal intensities is highly correlated, there are small satellite pulses at +/- 550fs. The unwanted replica pulses are a common difficulty with parametric shaping. Their origin is most likely attributable to imperfect transmission control. This allows some spectral components to be transmitted instead of being completely suppressed. One could imagine this situation being analogous to a half waveplate/polarizer combination, where the half waveplate is unable to completely convert P polarization to S polarization. Furthermore, the transmission of S polarization through the P polarizer is not completely suppressed due to the polarizer having less than 100 percent extinction of S polarization. Another contributing factor is the assumption of a perfectly Gaussian spectrum used in calculating the spectral filter function. The shape and smoothness of the shaped pulses spectrum is crucial for the generation of the proper temporal profiles. Unwanted modulations and a less than perfect Gaussian spectrum envelope, will result in unwanted and unavoidable additions to the retrieved temporal profiles. The phases of both the specified and measured temporal intensities are flat. The calculated and measured temporal separation of the intensity maxima are 400fs as required.

The measured spectral phases and amplitudes are shown in Fig. 2(d) and these correspond well to the required phase and spectral intensity shown in Fig. 2(b). The measured spectrum has the same number of amplitude oscillations compared to Fig. 2(b) and the amplitude of the fringes is also in agreement. The measured phase in Fig. 2(d) is also a periodic step function and the phase is flat across each of the spectral modulations.

To illustrate our ability to control more complex aspects of an ultrafast laser pulse shape, we designed temporal features that have ever increasing complexity.

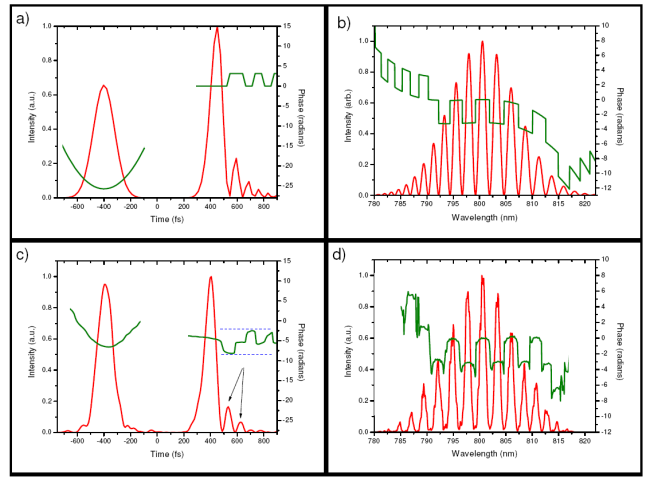

Fig. 3. a), The target pulse shape is temporal phase (green) and amplitude (red). b), The target pulse shape spectral phase (green) and amplitude (red) c), The measured temporal phase (green) and amplitude (red). d), The target pulse shapes spectral phase (green) and amplitude (red).

Figure 3(a) shows one example of a desired temporal pulse shape where the two pulses are separated in time by 800fs with a particular chirp on each pulse feature. We specify that one pulse should be linearly chirped by $-4\times10^3 fs^2$ and the other quadratically chirped by $4\times10^5 fs^3$.

The spectral intensity calculated by the SPECIFIC algorithm required to produce the temporal features is shown in Fig. 3(b). The required spectral intensity is again sinusoidal with a Gaussian envelope, and there are a series of phase steps. In this case the phase steps have a sigmoidal modulation and there is a higher complexity as can be seen. The spectral phase is no longer intuitive.

The temporal phase and amplitude corresponding to the pulse shape specified in Fig. 3(b) is shown in Fig. 3(c) as measured by SEA TADPOLE. Comparison of Fig. 3(a) with 3(c) reveals that the experimental pulses are in excellent agreement with the specified temporal phase, amplitude, and feature positions. The temporal positions of the intensity maxima are at their prescribed positions -400fs and +400fs. The shape of the recovered temporal amplitudes for the -400fs feature are in excellent agreement with the specified amplitude of Fig. 3(a). The measured temporal intensity spans from -600fs to -200fs in agreement with the specified pulse. The measured temporal amplitude at the 400fs position has the same intensity modulations which span into positive time, denoted by the black arrows in Fig. 3(c).

The temporal phases of the recovered pulses display the same characteristic contours for the specified spectral chirps. In Fig. 3(a), the temporal phase of the intensity maxima at the -400fs position is parabolic and positive. Indeed, the recovered temporal phase in Fig. 3(c) at -400fs is a positive parabola. There is strong agreement of the temporal phases for the pulses predicted at the 400fs position. As shown in Fig. 3(a), the intensity maxima at 400fs has a flat phase, with satellite pulses that extend into the positive temporal domain. The satellite pulses also have flat phases. Furthermore, these satellite pulses have π phase modulations. This is also the case for the recovered pulse and its satellites around the 400fs feature in Fig. 3(c). The main pulse's phase is flat and the satellite pulses have flat phases which alternate by π, as denoted by the blue dashes.

The complexity of the temporal shape of the pulse has increased; therefore intuition predicts that there should be a correspondingly higher degree of modulation in the spectral amplitude and phase. One can see that both the calculated (Fig. 3(b)) and measured (Fig. 3(d)) spectral components overlap with regard to the fringe spacing and the number of relevant spectral fringes. Furthermore, the measured spectral interference fringes have the correct amplitudes.

In an effort to push the limits of previous SEA TADPOLE [35] measurements, we have increased the complexity of our pulse shapes by generating a sequence of three pulses.

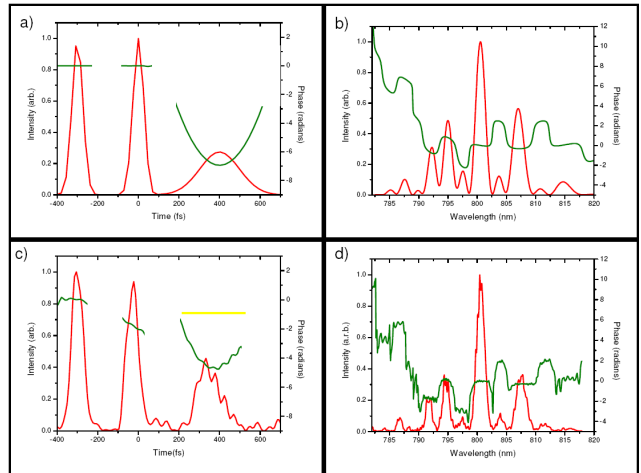

Figure 4. a), The target pulse shape's temporal phase (green) and amplitude (red). b), Recovered pulse from spectral components in d, the phase (red) and amplitude (green). c), Target pulse shapes spectral phase (green) and amplitude (red). d), The target pulse shapes spectral phase (green) and amplitude (red).

Figure 4(a) contains the theoretical temporal pulse profiles, which have temporal positions of -300 fs, 0 fs, and 400 fs. In addition, the pulse at 400fs has been linearly chirped by $5 \times 10^3 fs^2$. The SPECIFIC algorithm generates the spectral intensity and phase required for this sequence as shown in Fig. 4(b). In comparison with the previous calculations, both the spectral intensity and phase are non intuitive and complex.

The prescribed temporal phase of Fig. 4(a) is in good agreement with the phase measured in Fig. 4(c). The temporal positions of the features (Fig. 4(c)) are measured by SEA TADPOLE to occur the specified positions of -300 fs, 0 fs, and 400 fs. The shape of the measured temporal intensities, Fig. 4(c) at the -300fs and 0fs positions strongly coincide with the specified intensities in Fig. 4(a); and are Gaussian in shape. The intensities at the position

400fs is consistent with the specified pulse in Fig. 4(a), and is stretched temporally as specified.

The temporal phases of the recovered pulses are in good agreement with the structure specified in their specified counterparts. In Fig. 4(c) the temporal phase for the intensities at the -300fs and 0fs positions are virtually flat, mirroring the same composition as their corresponding equivalents of Fig. 4(a). The pulse at 400fs has a positive parabolic temporal phase (denoted by the yellow bar), as is expected from the linearly chirped pulse from Fig. 4(a).

The spectral phase and intensity components show a higher degree of modulation in comparison to the previous demonstrations. Both the calculated spectral intensity and phase shown in Fig. 4(b) and the measured spectral components shown in Fig. 4(d), are in accord with respect to the fringe spacing and the number of relevant interference fringes. Furthermore, the measured spectral interference fringes have the correct amplitudes. Correct spectral fringe spacing and amplitude will produce a more accurate recovered temporal pulse profile.

Fig. 4(b) and d have a complex spectral phase. We specified three pulses one of which has a linear chirp. The linear chirp requirement will be embodied in a quadratic spectral phase; however, beyond this the overall spectral phase contour is unintuitive. Despite the unintuitive spectral phase of Fig. 4(c), the measured spectral phase shown in Fig. 4(d) agrees very well with the calculated spectral phase.

Figure 5(a) details an experiment where three pulses are specified with temporal separations of -400fs, 0fs, and 400fs.

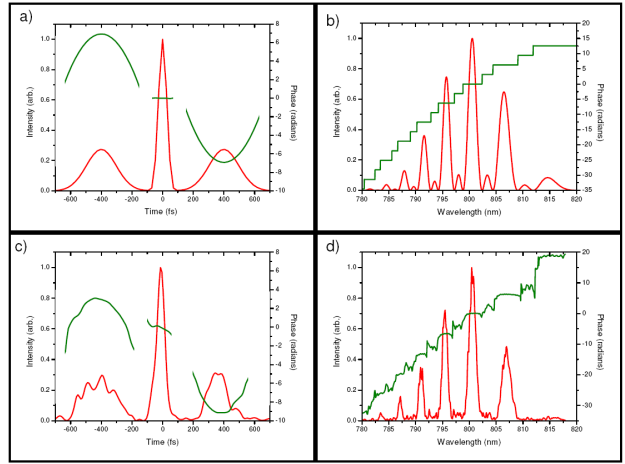

Fig. 5. a), Target pulse shape's temporal phase (green) and amplitude (red). b), Recovered pulse from spectral components in d, the phase (red) and amplitude (green). c) Target pulse shapes spectral phase (green) and amplitude (red). d), The target pulse shapes spectral phase (green) and amplitude (red).

The pulses at +/- 400fs have quadratic chirps valuing +/- 4000fs$^2$, respectively. The spectral intensity and phase modulations calculated by the SPECIFIC algorithm are shown in Fig. 5(b). Again these are non intuitive and complex.

The temporal phase and amplitude measured by SEA TADPOLE is shown in Fig. 5(c). The temporal profiles shown in Fig. 5(a) and c are similar. The temporal positions of the recovered intensities are at the positions stipulated by Fig. 5(a), -400 fs, 0 fs, and 400 fs. The recovered temporal intensity maxima for the -400fs, 0fs, and 400fs features in Fig. 5(c), are in good agreement with the calculated intensity profiles shown in Fig. 5(a). The intensities at the -400fs and 400fs features in Fig. 5(a) are both stretched in time, as expected for linearly chirped pulses. The intensity profile at the position 0fs of Fig. 5(a) is consistent with the measured pulse in Fig. 5(c), this feature is a near transform-limited pulse.

The temporal distributions for Fig. 5(a) and 5(c) reveal that the phase for the features at -400fs and 400fs exhibit a quadratic behavior. The parabolic phases in Fig. 5(a), are different by a minus sign, and the measured phases in 5(c) are in accord. The pulse in 5(a) at 0fs has a flat phase, as expected the recovered feature in 5(c) at 0 fs is effectively transform-limited as well.

As is shown in Fig. 5(b) and 5(d), the spectral modulations are also in good agreement; their spectral oscillations in intensity and phase have similar character. Fig. 5(b) and d have a complex spectral phase signature. The spectral phase for Fig. 5(b) and (d) are consistent in their step function character and their profiles superimpose as well.

As a final test of the SPECIFIC technique, we generated more complex three-pulse sequence.

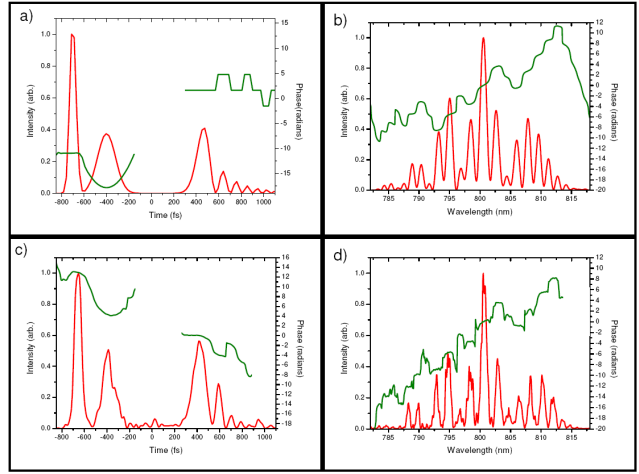

Fig. 6. a), Target pulse shapes temporal phase (green) and amplitude (red). b), Recovered pulse from spectral components in d, the phase (red) and amplitude (green). c), Target pulse shapes spectral phase (green) and amplitude (red). d), The target pulse shapes spectral phase (green) and amplitude (red).

In this sequence, the pulses are temporally positioned asymmetrically at delays -700fs, -400fs, and 400fs as shown in Fig. 6(a). The pulse's complexity is increased further with the additional requirement that the pulse at -400fs has a linear chirp of $4000fs^2$ and the pulse at 400fs has a quadratic chirp of $8x10^5 fs^3$ while the feature at -700 fs has flat phase. Fig. 6(b) shows the corresponding calculated spectral intensity and phase modulation returned from the SPECIFIC algorithm.

Figure 6(c) displays the temporal intensity and phase for the experimentally shaped pulse as measured by SEAPOLE. The positions of the temporal intensities for the -700 fs, -400 fs, and 400 fs features are shown in Fig. 6(a) and are accurately reflected in the SEA TADPOLE measurement shown in Fig. 6(c). The shape of the temporal intensity in 6(a) at -700fs is Gaussian, as is the measured profile in 6(c). The temporal feature in 6(a) at the position -400fs, is lengthened, as is the corresponding measured profile in 6(c). Furthermore, temporal profiles in both 6(a) and 6(c) extend in time space from ~ -600fs to -200fs. The temporal intensities at the position 400fs in 6(a) and 6(b) both have an intense feature followed by satellite modulations, characteristic of cubic spectral phase. The shape and intensities of these modulations correlate well between specified and measured pulses.

As for the temporal phase distributions, Fig. 6(a) displays a flat phase for the feature specified at -700fs, the corresponding measured feature in 6(c) has a flat phase as well. The feature at -400fs has a positive quadratic temporal phase in 6(a), the corresponding feature in 6(c) has a phase with a strong positive parabolic character. The temporal phase for the feature at 400fs in 6(a) has a flat phase for the main pulse and satellite pulses. The absolute phase also exhibits $\pi$ jumps. The phase for the measured pulse shape in 6(c) demonstrates the same satellite pulses and phase character for the measured feature at 400fs.

Figure 6(b) and d display the calculated and measured spectral amplitudes. As is evident from b and d, the fringe spacing and amplitudes for the specified and measured pulses coincide. The sub-pulses in Fig. 6(b) and 6(d) have quadratic and cubic spectral phases, respectively. The spectral phase for Fig. 6(b) and 6(d) are consistent in their profiles. Again neither the spectral intensity nor phase is intuitive for the desired shaped pulse. The ability to accurately measure these profiles suggests that we are constructing time dependent electric fields with high fidelity.

## 3. Conclusions:

This paper demonstrates that the SPECIFIC method can produce, apply, and measure complicated pulse shapes in the ultrafast regime. The recovered temporal phases and intensities from SEA TADPOLE are excellent evidence that the SPECIFIC algorithm can control the fine details of the temporal pulse profile. Furthermore, our open loop method further validates the ability of SEA TADPOLE to characterize and verify complex pulse shapes. The SPECIFIC method will be useful in the fields of coherent control, high harmonic generation, and optical metrology. In these fields there is a huge demand for a technique that can produce a complex pulse and simultaneously verify the pulse shape interacting with the physical system. In this way we anticipate measuring, mining, and enhancing physical insight for a multitude of pulses interacting with molecular systems.

## Acknowledgements:

This work was supported by grants from the National Science Foundation CHE No. 331390111 (R.J.L), Defense Advanced Research Projects Agency No. 311390111 (R.J.L), the Army Research Office No. 311390121 (R.J.L) and an STTR grant as managed by the Army Research Office (R.J.L), the Swiss NCCR (Weber), NSF SBIR grant #053-9595 (Trebino) and NSF fellowship IGERT-0221600 (Bowlan).